\documentclass[12pt]{article}
\usepackage[T1]{fontenc}
\usepackage[utf8]{inputenc}
\usepackage[margin=1in]{geometry}
\usepackage{amsmath,amssymb}
\usepackage{bm}
\usepackage{graphicx}
\usepackage{xcolor}
\usepackage{setspace}
\usepackage[super,sort&compress]{natbib}

\usepackage[hidelinks]{hyperref}
\raggedright
\title{Limits from nonlinear loss of spin-orbit-torque control of magnon transport}
\author{\parbox{\dimexpr\textwidth-2\tabcolsep\relax}{\centering
Lerato Takana\textsuperscript{1,3},
Katya Mikhailova\textsuperscript{1,3},
Charles Zheng\textsuperscript{1},
Sauviz Alaei\textsuperscript{2},
Xiangcheng Liu\textsuperscript{4},
Junwei Tong\textsuperscript{4},
Kwangyul Hu\textsuperscript{5},
Michael Flatté\textsuperscript{5},
Xiaoqin Li\textsuperscript{4}, and 
Yuri Suzuki\textsuperscript{1,3}\\[1ex]
\small
\textsuperscript{1}Department of Applied Physics, Stanford University, Stanford, CA, 94305, USA\\
\textsuperscript{2}Department of Physics, Stanford University, Stanford, CA, 94305, USA\\
\textsuperscript{3}SIMES, SLAC National Accelerator Laboratory, Menlo Park CA, 94025, USA\\
\textsuperscript{4}Department of Physics, The University of Texas at Austin, Austin, TX 78712, USA\\
\textsuperscript{5}Department of Physics and Astronomy, University of Iowa, Iowa City, Iowa 52242, USA
}}
\date{}
\begin{document}
\maketitle

\begin{abstract}
Controlling magnon currents is central to the development of magnonic devices, which exploit spin excitations for information transport without charge flow. Here, we use the spin-orbit torque (SOT) from a Pt gate to control the effective magnon damping in three-terminal devices fabricated from low-damping lithium aluminum ferrite (LAFO) thin films and identify the nonlinear damping that ultimately limits the gate response. With increasing gate current, the nonlocal magnon signal initially increases, but saturates and subsequently decreases above a critical gate current density. Brillouin light scattering microscopy reveals that this turnover coincides with the emergence of an SOT-driven auto-oscillation mode localized beneath the gate, whose frequency continuously redshifts with increasing gate current, characteristic of large-amplitude nonlinear dynamics, and with a suppression of the propagating magnon population away from the gate. Micromagnetic simulations reproduce the observed nonmonotonic gate response only when an amplitude-dependent effective damping is included, indicating that nonlinear losses, rather than saturation of spin injection, limit the degree of magnon modulation. Our results show that SOT compensation of magnon damping is ultimately limited by nonlinear damping attributed to auto-oscillations, providing a framework for understanding the limits of magnon current modulation in magnetic insulators.

\end{abstract}

\section*{Introduction}
Magnons, the collective spin excitations of magnetically ordered materials, can carry spin angular momentum through electrical insulators without charge transport, making them attractive information carriers for low-dissipation information processing and signal transmission \cite{hoffmann2007pure,Jansen2012SiliconSpintronics,Coll2019TowardsOxideElectronics,Ahopelto2019NanoElectronicsRoadmap}. Active control of magnon transport using spin currents offers a route toward reconfigurable magnonic devices, including functional interconnects and logic components. Realizing these applications requires efficient control of magnon currents and, importantly, an understanding of the physical mechanisms that ultimately limit this control. In recent years, magnon currents have been modulated and switched using spin currents generated through the spin Hall effect \cite{zheng2023,cornelissen2018three,wimmer2018spin,guckelhorn2021magnon,fulara2019spin,collet2016generation,Hamadeh2014SpinWaveDamping,Evelt2016HighEfficiency,Evelt2018CoherentPropagatingMagnons,Bai2024FieldFreeYIG,Bai2025FullScaleYIG,Tian2020YIGReorientation}. In a three-terminal nonlocal geometry, a magnon current is generated at a source contact and detected at a drain contact, while a gate electrode positioned between them applies a spin-orbit torque (SOT) to the magnetic insulator. The gate SOT modifies the local magnon population, thereby controlling the magnon current reaching the drain \cite{cornelissen2018three,wimmer2018spin,guckelhorn2021magnon,fulara2019spin}.

The SOT can enhance the magnon population by exerting an antidamping-like torque on the magnetization. In this picture, the gate current acts as a control knob for the effective magnon damping beneath the gate. As the gate current is increased, the SOT progressively compensates the intrinsic magnetic damping and can eventually drive the system into a nonlinear dynamical regime. At a critical current, complete damping compensation can lead to the onset of coherent auto-oscillations \cite{collet2016generation,Demidov2012MagneticNanoOscillations,Tsoi1998ExcitationSpinWaves, Ren2025LAFOAutoOscillations,Ren2023HybridSHNO}, while strong excitation of incoherent thermal magnon modes can also contribute to the enhanced magnon population. Previous experiments in yttrium iron garnet have demonstrated approximately a factor-of-two enhancement of the magnon current in this regime \cite{wimmer2018spin,guckelhorn2021magnon}. These observations raise an important question for magnon current control: what limits the magnon population and transport response when the applied SOT approaches and exceeds the damping-compensation regime? Understanding the evolution of the excited magnon modes and the associated nonlinear losses are therefore essential for ultimate control of magnon currents.

In this paper, we find that nonlinear damping attributed to auto-oscillations limits magnon current control in three-terminal Li$_{0.5}$Al$_{1.0}$Fe$_{1.5}$O$_{4.0}$ (LAFO) spintronic devices. We observe a nonmonotonic gate dependence of the source-drain magnon signal: the signal initially increases with increasing dc current in the Pt gate, then saturates, and finally decreases above a critical gate current density. Brillouin light scattering (BLS) microscopy, which probes the magnon population both beneath the gate and in the channel next to it, reveals the emergence of an SOT-driven auto-oscillation mode whose frequency continuously redshifts with increasing gate current, characteristic of large-amplitude nonlinear dynamics. At higher gate currents, the BLS signal from propagating magnons is suppressed along with the localized auto-oscillation mode. Micromagnetic simulations reproduce the observed nonmonotonic gate response only when a current-dependent effective damping is included, indicating that nonlinear damping limits the increase of the magnon population and, consequently, the degree of modulation. Our results demonstrate that the SOT cannot increase source-drain magnon signal arbitrarily: the attainable modulation is therefore set not by the efficiency of spin injection but by the onset of nonlinear damping, identifying the suppression of nonlinear losses, rather than stronger spin injection, as the key to more effective magnon current control.

\section*{Results}
\subsection*{Nonlocal measurements}
\begin{figure}[tbp]
    \centering
    {\includegraphics[width=\linewidth,height=0.65\textheight,keepaspectratio]{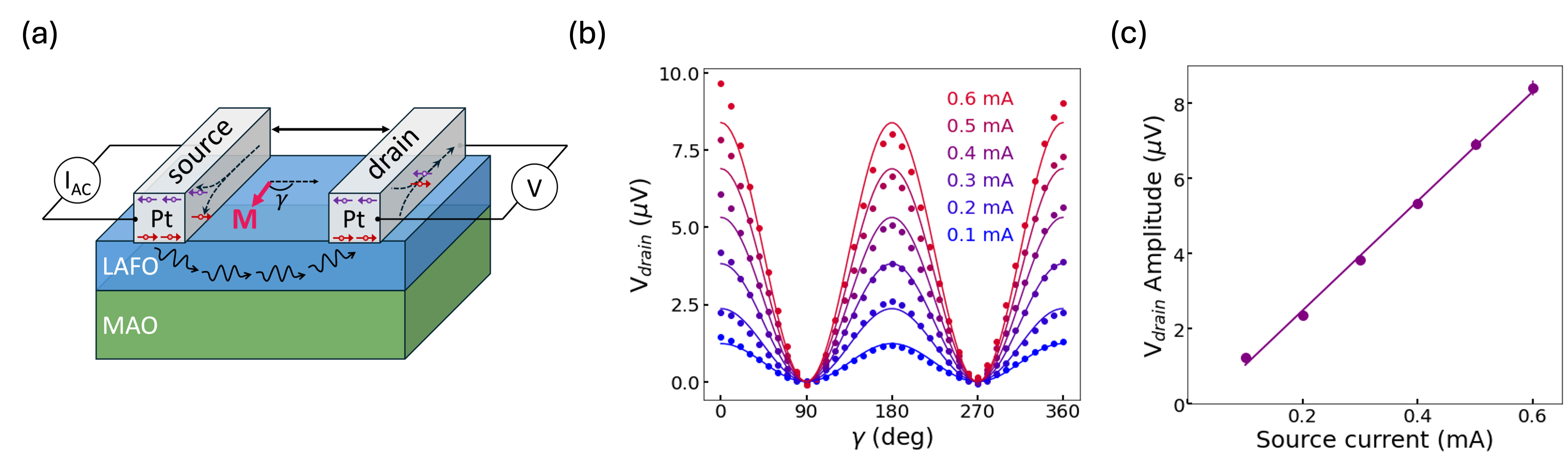}}
    \caption{\textbf{Nonlocal measurements of a two-terminal device showing magnon current generation.} (a) Schematic of the nonlocal measurement geometry. An ac current (f = 9.8 Hz) is applied to the Pt source, while the first-harmonic voltage is measured at the Pt drain using a lock-in amplifier. The source and drain are separated by an edge-to-edge distance of (d = 300 nm). Pt nanowires are 4 nm thick, 300 nm wide and 40 $\mu$m long. An in-plane magnetic field of $\mu_0$H = 80 mT is applied to saturate the magnetization. Measurements were performed at 300 K. (b) Angular dependence of the first-harmonic drain voltage measured at different ac source currents. The in-plane angle, $\gamma$, is defined as the angle between the spin polarization generated in the Pt source and the magnetization of the LAFO film. Solid lines are fits to (V = V$_{drain}$cos$^2(\gamma$)). (c) Extracted voltage amplitude, V$_{drain}$, as a function of the source current. The linear dependence of V$_{drain}$ on the AC current is consistent with electrically generated magnon transport.}
    \label{fig:1}
\end{figure}

\begin{figure}[tbp]
    \centering
\includegraphics[width=\linewidth,height=0.65\textheight,keepaspectratio]{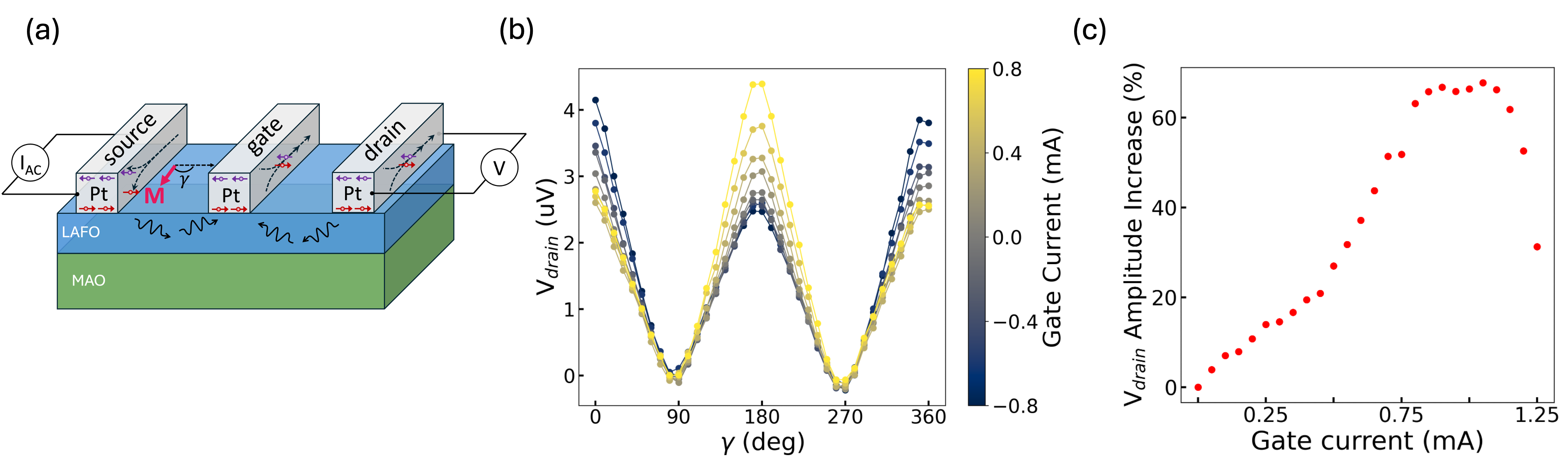}
    \caption{\textbf{Nonlocal measurements of a three-terminal device for magnon current control.} (a) Schematic of the three-terminal nonlocal device used for magnon control in Li$_{0.5}$Al$_{1.0}$Fe$_{1.5}$O$_4$. Three Pt nanowires (4 nm thick, 300 nm wide, and 40 $\mu$m long) are separated by an edge-to-edge distance of 300nm. With an ac current (f = 9.8 Hz) applied to the source and a dc current applied to the gate, the first-harmonic nonlocal voltage is measured at the drain using a lock-in amplifier. An in-plane magnetic field of $\mu_0$H = 80 mT saturates the magnetization. All measurements were performed at 300 K. (b) Angular dependence of the first-harmonic non-local voltage for different dc gate currents at a fixed I$_{\mathrm{AC}}$ = 0.6 $\mathrm{mA}$. The angle $\gamma$, defined in the schematic, is between the spin accumulation generated at the Pt/LAFO interface and the magnetization of the LAFO film. Solid lines are guides to the eye. (c) Percentage increase of the drain voltage amplitude relative to its value at zero gate current as a function of dc gate current for ($\gamma = 180^\circ$), where the spin accumulation is parallel to the magnetization. The non-local voltage increases with increasing gate current at low bias and is suppressed above a critical current density.}
    \label{fig:2}
\end{figure}

To investigate electrical control of magnon transport in LAFO, we performed nonlocal transport measurements in three-terminal devices comprised of a source, a drain and an electrical gate in between (Fig.~\ref{fig:2}(a)). As a control we also fabricated two terminal devices with just a source and drain (Fig.~\ref{fig:1}(a)). In the two-terminal geometry, an alternating charge current is applied to the Pt source, where the spin Hall effect generates a transverse spin current resulting in spin accumulation at the Pt/LAFO interface. The interfacial spin accumulation transfers angular momentum to the magnetic moments in LAFO, creating a non-equilibrium magnon population that diffuses through the magnetic insulator \cite{hoffmann2007pure,sinova2015spin,maekawa2013spin}. The ac current also generates a temperature gradient and, in turn, magnons due to the spin Seebeck effect (SSE) \cite{uchida2010spin,xiao2010theory,Giles2015,cornelissen2015long,Uchida2008,Uchida2014}. At the drain a distance d away, electrically and thermally generated magnons are detected via the inverse spin Hall effect as the first and second harmonic detector voltage signals, respectively \cite{hoffmann2007pure,maekawa2013spin}.

In the case of electrically generated magnons, efficiency of both magnon injection and detection depends on the projection of the spin accumulation at the Pt/LAFO interface onto the magnetization direction of LAFO. Consequently, under a saturating in-plane magnetic field, the nonlocal signal is expected to vary as V = V$_{drain}$cos$^{2}\gamma$,
where  V$_{drain}$ is proportional to the injected magnon current and $\gamma$ is the angle between the spin polarization generated by the spin Hall effect and the magnetization of LAFO. The two cos$\gamma$ factors arise from the injection and detection efficiencies, respectively \cite{cornelissen2015long}. Figure~\ref{fig:1}(b) shows the measured angular dependence together with a fit to V = V$_{drain}$cos$^{2}\gamma$ for the two terminal geometry. The extracted amplitude (V$_{drain}$), shown in Fig.~\ref{fig:1}(c), scales linearly with the applied ac current as expected.

To actively control magnon transport, we patterned devices with a third Pt nanowire between the source and drain to serve as a magnon control gate (Fig.~\ref{fig:2}(a)). Passing a dc current through the gate generates a spin accumulation at the Pt/LAFO gate interface via the spin Hall effect, thus allowing the local magnon population under the gate to be acted on by a spin-orbit torque \cite{cornelissen2018three}. Depending on its sign, this torque either counteracts (antidamping) or adds to the intrinsic magnetic damping. As shown in Fig.~\ref{fig:2}(b), the nonlocal voltage is enhanced when the spin accumulation in the gate is parallel to the LAFO magnetization and suppressed when it is antiparallel to the LAFO magnetization. For instance, when $\gamma = 180^{\circ}$, positive and negative dc currents in the gate lead to an enhancement and suppression of the drain voltage, respectively. When the magnetization of LAFO is reversed, $\gamma = 0^{\circ}$, we observe the opposite behavior such that positive dc currents show a suppression and negative dc currents show an enhancement.

The nonlocal drain signal exhibits a pronounced nonlinear dependence on the gate current (Fig.~\ref{fig:2}(c)). We compare the drain voltage increase with its value when there is zero gate current. At low current densities, the drain voltage increases linearly with dc current, consistent with the expected linear response of spin-orbit-torque-driven magnon injection, with a gate efficiency of 71\%/mA. The modulation reaches a maximum of 51\% at a gate current of $\approx0.7~\mathrm{mA}$, corresponding to a current density of $\approx5.8\times10^{11}~\mathrm{A/m^2}$ in the Pt gate. The low saturation magnetization and low damping of LAFO reduce the torque required to compensate the magnon damping, enabling this modulation at relatively low current density \cite{zheng2023, Omahoney2023,Takana2025LowDampingLAFO,Alaei2025EfficientSpinTransport,tong2026direct,pal2026low,Li2026NearlyIsotropic}. As the current density is further increased, the response deviates from linearity, indicating the onset of strong nonlinear spin dynamics. At a critical current, the nonlocal signal begins to decrease despite increasing anti-damping torque, suggesting that additional nonlinear damping and magnon-magnon scattering processes limit the magnon population. To determine how the gate modifies the magnon damping and population in this high-current regime, we complement transport measurements with Brillouin light scattering (BLS) measurements and micromagnetic simulations.

\subsection*{Brillouin light scattering}
\begin{figure}[tbp]
    \centering
\includegraphics[width=\linewidth,height=0.65\textheight,keepaspectratio]{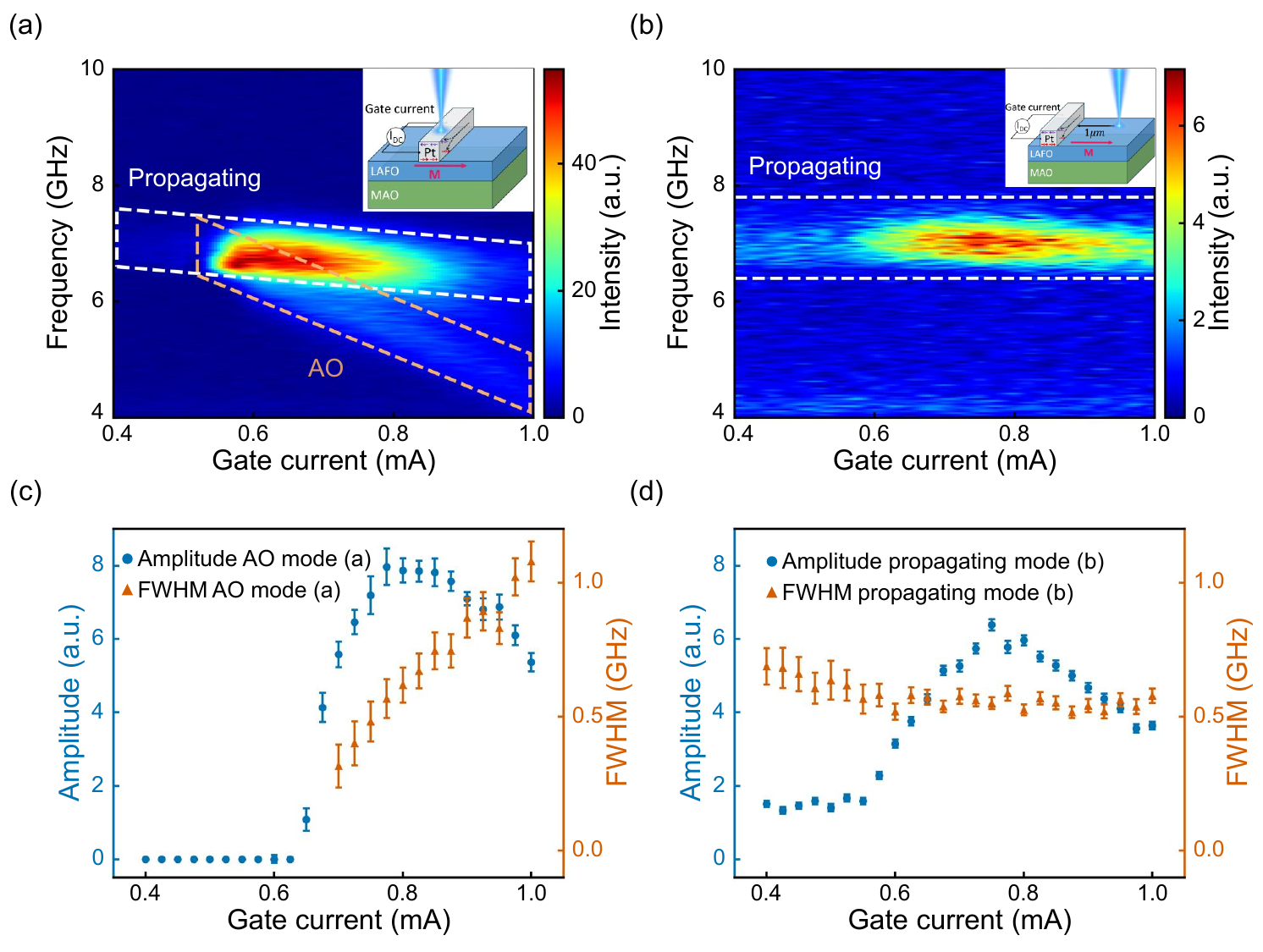}
    \caption{\textbf{Brillouin Light Scattering measurements on a Pt/LAFO bilayer.} (a) BLS spectra measured with the laser spot positioned on top of the Pt bar under DC currents ranging from \(0.4\) to \(1.0~\mathrm{mA}\) in \(0.025~\mathrm{mA}\) increments. (b) Same as (a) but measured 1 $\mu$m from the Pt bar. (c) Current dependence of the fitted amplitude and full width at half maximum (FWHM) linewidth of the auto-oscillation mode extracted from the spectra in (a). (d) Current dependence of the fitted amplitude and full width at half maximum (FWHM) linewidth of the propagating magnon mode extracted from the spectral map in (b).}
    \label{fig:BLS}
\end{figure}

To identify the magnon modes responsible for the saturation and subsequent decrease of the drain voltage, we performed BLS microscopy as a function of the dc gate current, with the laser spot positioned either on the Pt gate or $1~\mu\mathrm{m}$ away from it [Figs.~\ref{fig:BLS}(a) and (b)]. At all currents, the spectra contain a dominant mode near $7~\mathrm{GHz}$. Because BLS detects magnons with a small but finite wave vector, we attribute this mode to the low-$k$ propagating magnons that carry spin angular momentum from the gate region toward the drain.

When the gate current exceeds approximately $0.6~\mathrm{mA}$, an additional lower-frequency mode emerges in the spectra measured at the gate (Fig.~\ref{fig:BLS}(a)). We assign this mode to an SOT-driven auto-oscillation (AO) mode, which appears once the antidamping torque compensates the effective magnon damping \cite{bender2014dynamic,Takei2019}. Its frequency redshifts continuously with increasing current (Fig.~\ref{fig:BLS}(c)), consistent with the nonlinear frequency shift of large-amplitude auto-oscillations in films with in-plane magnetization \cite{fulara2019spin,Slavin2009Nonlinear}. Because the AO frequency lies below the propagating magnon band, the mode cannot propagate and remains localized beneath the gate; accordingly, it is absent in the spectra measured $1~\mu\mathrm{m}$ away (Fig.~\ref{fig:BLS}(b)) and contributes negligibly to the nonlocal signal. The AO intensity reaches a maximum near $0.8~\mathrm{mA}$ and then decreases, while its linewidth broadens with increasing current [Fig.~\ref{fig:BLS}(c)], indicating enhanced dissipation associated with nonlinear damping \cite{divinskiy2019controlled}.

The propagating magnon mode measured $1~\mu\mathrm{m}$ from the gate directly reflects the population that reaches the drain [Fig.~\ref{fig:BLS}(d)]. Its amplitude increases with gate current but decreases above approximately $0.75~\mathrm{mA}$, mirroring the turnover of the drain voltage in Fig.~\ref{fig:2}(c), and its linewidth remains narrower than that of the AO mode throughout. Together, these observations indicate that once the AO mode forms, additional SOT no longer reduces the effective damping of the propagating magnons. Instead, energy supplied by the gate is diverted into the localized AO mode and dissipated through nonlinear damping, limiting the propagating magnon population at high currents.

\subsection*{Micromagnetic simulations}

\begin{figure}[tbp]
    \centering
  \includegraphics[width=0.85\linewidth,height=0.65\textheight,keepaspectratio]{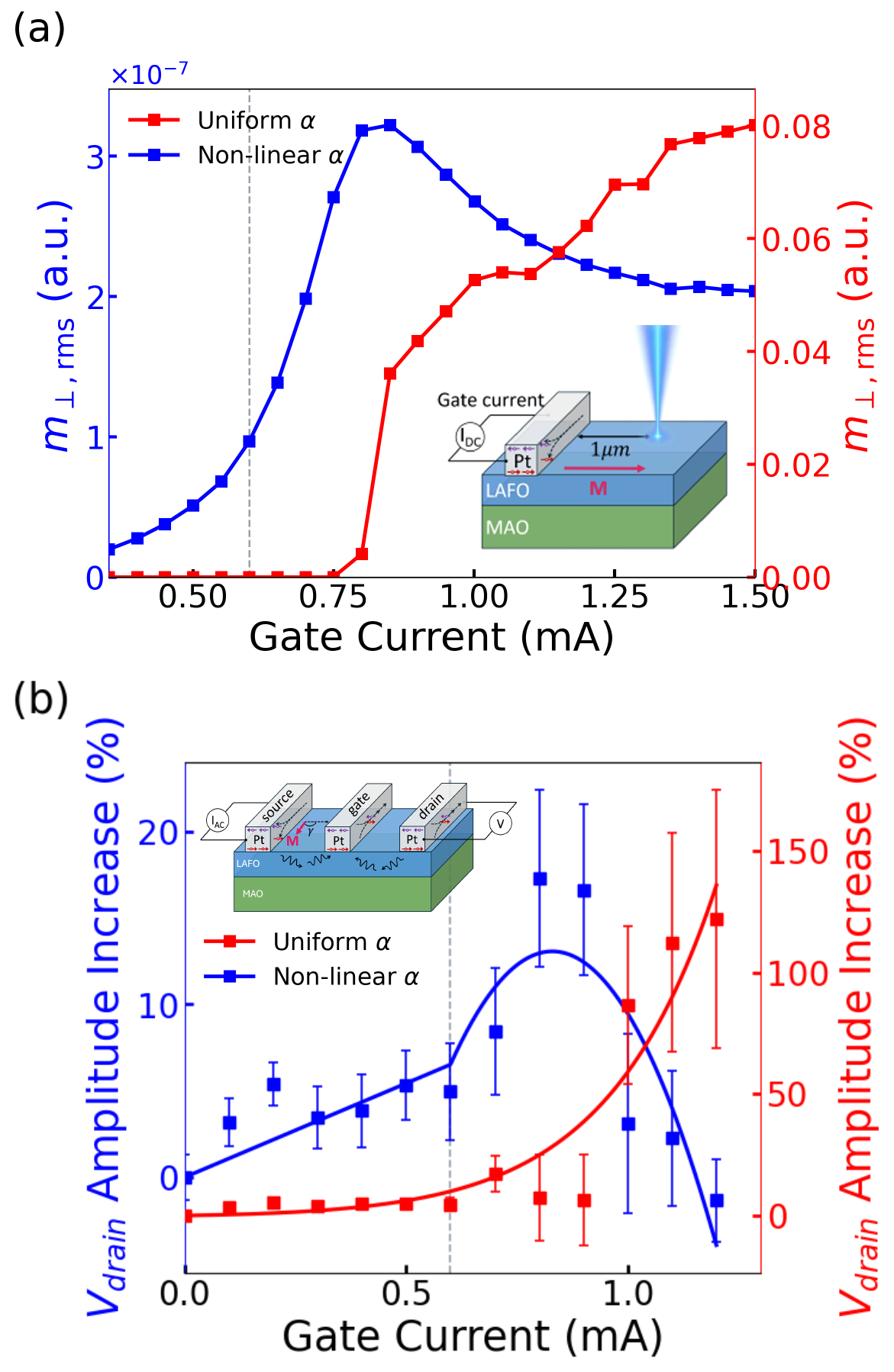}
  \caption{\textbf{Micromagnetic simulations of the nonlocal and BLS measurements.} Simulated (a) $m_{\perp,rms}$ of the propagating magnons and (b) $V_{drain}$ amplitude increase as functions of the DC gate current ($I_g$) for the current independent damping (Uniform $\alpha$, red) and the phenomenological current-dependent damping (Non-linear $\alpha$, blue). The gray dashed line marks the threshold current, $I_{th}=0.6~\text{mA}$. The solid lines in (a) are a guide to the eye and (b) red and blue are fits to an exponential function and two piecewise functions with linear and quadratic terms, respectively. \label{Fig_Sim}}
\end{figure}

To determine whether nonlinear damping can account for the observed turnover, we performed micromagnetic simulations of both the propagating magnon amplitude and the nonlocal signal. We first considered a current-independent Gilbert damping $\alpha$ and calculated the root-mean-square (rms) transverse magnetization precession amplitude ($m_{\perp,rms}$) of the propagating magnons $1~\mu\mathrm{m}$ from the gate (Methods). The corresponding simulation results are shown by the red curve in Fig.~\ref{Fig_Sim} (a). The constant damping simulation produced only a weak change below $0.75~\text{mA}$, followed by a monotonic increase above $0.75~\text{mA}$. It did not reproduce the experimentally observed reduction at large gate currents in Fig.~\ref{fig:BLS} (d).

To test whether additional high-current dissipation could produce the turnover, we introduced a phenomenological current-dependent correction to the effective Gilbert damping. This choice is motivated by the current-dependent linewidth of the auto-oscillation mode in [Figure~\ref{fig:BLS}(c)], indicating the presence of nonlinear damping\cite{divinskiy2019controlled}. Since nonlinear losses are amplitude dependent \cite{Slavin2009Nonlinear,Lee2022Nonlinear}, we model their current dependence above $I_{th}$ through the spatially averaged effective damping:
\begin{equation}
    \alpha_{eff}\left(I_g\right)=\begin{cases}
\begin{array}{c}
\alpha_0,\\
\alpha_0+\alpha_1\zeta+\alpha_2\zeta^2,
\end{array} & \begin{array}{c}
I_{g}\leq I_{th},\\
I_{g}>I_{th}.
\end{array}\end{cases}\label{nldamp_2}
\end{equation}
Here, $I_g$ is the gate current, $\zeta=I_g-I_{th}$, $\alpha_0=0.0015$ is the baseline damping, and $\alpha_1$ and $\alpha_2$ are the linear and quadratic coefficients, respectively. Note that this expression phenomenologically represents amplitude-dependent losses, not an intrinsic current dependence of the Gilbert damping.

A linear correction was unable to reproduce the high-current suppression, while finite $\alpha_1$ tended to obscure the required quadratic curvature. Therefore, $\alpha_1=0$ and $\alpha_2=6\times10^3~\text{A}^{-2}$ were considered in the simulations. The resulting $m_{\perp, rms}$ shown by the blue curve in Fig.~\ref{Fig_Sim} (a) reproduces the non-monotonic current dependence observed experimentally and agrees well with the BLS measurement of the propagating magnon mode in Fig.~\ref{fig:BLS} (d).

Next, we examined the effect of the nonlinear damping in the nonlocal magnon transport device. Thermal fluctuations and nonlinear dynamics above $I_{th}$ introduced substantial scatter, so the responses were fitted to extract their current dependence (Supplementary Information, Section II). For the current dependent damping case, separate fits were applied over $0\leq I_g <0.6~\text{mA}$, and $0.6\leq I_g <1.2~\text{mA}$ whereas the uniform damping response was fitted with an exponential function (Supplementary Information, Section II). As shown in Fig.~\ref{Fig_Sim} (b), a nonmonotonic lock-in-like response appears only for the nonlinear damping case. The signal initially increases with the gate current, turns over at $I_g\approx0.9~\text{mA}$ and decreases at larger gate currents. These results demonstrate that enhanced high-current dissipation reproduces the qualitative turnover observed in Fig.~\ref{fig:2} (c).

\section*{Discussion}
Our combined transport, BLS, and micromagnetic results reveal that the same spin-orbit torque responsible for enhancing the propagating magnon population also drives the system toward a nonlinear dynamical regime. With increasing gate current, the antidamping torque progressively compensates the intrinsic damping and eventually drives the emergence of a localized auto-oscillation mode beneath the gate. Above this threshold, the additional energy supplied by the gate does not translate into a continued increase of the propagating magnon population. Instead, an increasing fraction of the excitation is associated with the localized auto-oscillation and nonlinear magnon-magnon scattering, providing an additional dissipation channel that limits the magnon signal reaching the drain \cite{collet2016generation,Slavin2009Nonlinear,Lee2022Nonlinear}. The continuous redshift of the auto-oscillation frequency with increasing gate current is consistent with the large-amplitude nonlinear evolution of this mode, while the ability of the amplitude-dependent damping model to reproduce the observed turnover indicates that nonlinear damping plays a central role in limiting the modulation. Thus, the nonmonotonic gate response is not simply a saturation of spin injection; rather, it emerges from the redistribution and dissipation of the injected energy through nonlinear magnetic dynamics. The onset of auto-oscillation therefore defines the boundary of continuous damping control in this geometry. Engineering this boundary, for example through the intrinsic damping, device geometry, mode confinement, or magnetic anisotropy, provides a route toward larger modulation depths \cite{Dvornik2018,Lee2022Nonlinear,fulara2019spin}.

\section*{Conclusion}

In summary, we demonstrate direct evidence of spin-orbit-torque-driven auto-oscillation modes following the compensation of magnetic damping in a magnetic insulator. The emergence of these auto-oscillations introduces an additional nonlinear damping channel. More specifically, nonlocal electrical measurements show that the source-drain magnon signal initially increases with increasing dc current in the Pt gate, but saturates and subsequently decreases above a critical gate current density. BLS measurements reveal that this nonmonotonic response coincides with the emergence of a localized spin-orbit-torque-driven auto-oscillation mode  and a reduction of the propagating magnon population. The continuous evolution of this mode with increasing gate current, together with micromagnetic simulations incorporating current-dependent effective damping, indicates that nonlinear losses associated with the auto-oscillatory state limit the attainable magnon modulation. Our results establish nonlinear damping, rather than spin injection, as the fundamental bottleneck of electrically controlled magnon transport, and identify the suppression of nonlinear losses near the auto-oscillation threshold as the critical step toward efficient magnon control in magnetic insulators.

\section*{Methods}

\textit{Synthesis. --- }  Epitaxial thin films of Li$_{0.5}$Al$_{1}$Fe$_{1.5}$O$_{4}$ (LAFO) were grown on single-crystalline MgAl$_{2}$O$_{4}$ (001) using pulsed laser deposition at substrate temperature of 425$^\circ$C in 15 mTorr of O$_{2}$. A pressed target of Li$_{0.6}$Al$_{1}$Fe$_{1.5}$O$_{4}$ (LAFO) from Toshiba Corporation was pre-ablated with 248 nm KrFl laser at a laser fluence of 2.8 J/cm$^2$ and a repetition rate of 2 Hz. The excess Li compensates for the high volatility of Li under high temperature growth. At the end of the deposition, the films were cooled down in 150 Torr of O$_{2}$ to room temperature. 4 nm of Pt (300 nm wide) was dc sputtered at 200 W in 3 mTorr of Ar at room temperature on top of LAFO films.

%\textit{Characterization. --- } X-ray diffraction measurements were performed on an Panalytical Empyrean 2 diffractometer, showing excellent crystallinty and epitaxy. X-ray reflectivity measurements were performed to confirm film thicknesses. Broadband ferromagnetic resonance was performed in a co-planar waveguide geometry with a 10 dBm RF source to probe the magnetic loss on the order of $\alpha$$\approx$ 1.5 x 10$^{-3}$ in the the spatially uniform $k=0$ mode. Magnetic hysteresis loops were measured using a Quantuam Design Evercool SQUID magnetometer and found to have in-plane magnetic anisotropy.

\textit{Device Fabrication and Electrical Measurements. --- } Devices comprised of two and three 4 nm thick Pt electrodes on LAFO films were defined by electron beam lithography followed by a lift-off process (Fig.~\ref{fig:1}(a)).  AuPd electrical contacts were sputtered on the Pt electrodes in 3 mTorr of Ar after patterning with photolithography using a Durham Magneto Optics MicroWriter ML3 direct-write system. The edge-to-edge spacing, d, between the 300 nm wide Pt nanowires is 300 nm for all devices. Nonlocal voltage at the drain was measured using an SR-830 lock-in amplifier and the current at the source and gate were from a Keithley 6221 current source.

\textit{Brillouin light scattering. --- }
BLS measurements were performed in a backscattering geometry using a tandem Fabry--Pérot interferometer (TFP-1, JRS Scientific Instruments). A \(488~\mathrm{nm}\) laser with a power of \(2~\mathrm{mW}\) was focused through a \(100\times\) objective to a spot approximately \(1~\mu\mathrm{m}\) in diameter. The laser spot was positioned on top of the Pt bar to probe the local magnon population beneath the gate, or $1~\mu\mathrm{m}$ away from it to probe the propagating magnons.

\textit{Micromagnetic simulations. --- } Micromagnetic simulations were performed using MuMax3 \cite{Vansteenkiste2014MuMax}. To compare with the BLS experiment, we calculated $m_{\perp,rms}$ by averaging over a $1.5~\mu\text{m}$ region of LAFO film located $1~\mu\text{m}$ away from the gate. The simulations were performed at $T=0~\text{K}$, and the magnetization dynamics were band-pass filtered over the frequency range of $6-8~\text{GHz}$ prior to calculating $m_{\perp,rms}$, isolating the propagating magnon band. Next, we simulated the three-terminal nonlocal device. A quantity proportional to the spin-pumping induced inverse spin-Hall voltage was calculated from the magnetization dynamics in the magnetic region beneath the drain \cite{Tserkovnyak2002SpinPumping1,Tserkovnyak2002SpinPumping2}: $V\propto\left<\left[\mathbf{m}\times\dot{\mathbf{m}}\right]_y\right>_{t,\Omega_{D}}$. Here, the brackets denote averaging over time and over the magnetic volume $\Omega_{D}$ and the reduced magnetization $\mathbf{m}=\mathbf{M}/M_s$. The $y$-component was selected according to the experimental voltage detection geometry. To approximate the bipolar ac excitation used in the lock-in measurement, simulations were performed for the source current $I_s=+0.6$ and $-0.6~\text{mA}$. The lock-in-like response was then calculated as
\begin{equation}
    V_{lock}\left(I_g;I_s\right)=\dfrac{V\left(I_g;+0.6~\text{mA}\right)-V\left(I_g;-0.6~\text{mA}\right)}{2}
\end{equation}
Thermal fluctuations were included using an effective simulation temperature $T_{eff}=30~\text{K}$. This reduced temperature was chosen to suppress the thermal fluctuations. The reduced temperature lowers the thermal magnon population and thus underestimates the signal amplitude. The simulations are therefore intended to capture qualitative trends rather than absolute magnitude.

\clearpage % Flush figures/tables before the bibliography.
% DRAFT: ref.bib and naturemag.bst must be present in this project.
% FINAL SUBMISSION: replace BOTH commands below with the complete contents
% of main.bbl, including its thebibliography environment.
% Generate this .bbl with BibTeX, not Biber/biblatex.
\bibliographystyle{naturemag}
\bibliography{ref}

\begin{thebibliography}{10}
\expandafter\ifx\csname url\endcsname\relax
  \def\url#1{\texttt{#1}}\fi
\expandafter\ifx\csname urlprefix\endcsname\relax\def\urlprefix{URL }\fi
\providecommand{\bibinfo}[2]{#2}
\providecommand{\eprint}[2][]{\url{#2}}

\bibitem{hoffmann2007pure}
\bibinfo{author}{Hoffmann, A.}
\newblock \bibinfo{title}{Pure spin-currents}.
\newblock \emph{\bibinfo{journal}{physica status solidi c}} \textbf{\bibinfo{volume}{4}}, \bibinfo{pages}{4236--4241} (\bibinfo{year}{2007}).

\bibitem{Jansen2012SiliconSpintronics}
\bibinfo{author}{Jansen, R.}
\newblock \bibinfo{title}{Silicon spintronics}.
\newblock \emph{\bibinfo{journal}{Nature Materials}} \textbf{\bibinfo{volume}{11}}, \bibinfo{pages}{400--408} (\bibinfo{year}{2012}).

\bibitem{Coll2019TowardsOxideElectronics}
\bibinfo{author}{Coll, M.} \emph{et~al.}
\newblock \bibinfo{title}{Towards oxide electronics: a roadmap}.
\newblock \emph{\bibinfo{journal}{Applied Surface Science}} \textbf{\bibinfo{volume}{482}}, \bibinfo{pages}{1--93} (\bibinfo{year}{2019}).

\bibitem{Ahopelto2019NanoElectronicsRoadmap}
\bibinfo{author}{Ahopelto, J.} \emph{et~al.}
\newblock \bibinfo{title}{Nanoelectronics roadmap for europe: From nanodevices and innovative materials to system integration}.
\newblock \emph{\bibinfo{journal}{Solid-State Electronics}} \textbf{\bibinfo{volume}{155}}, \bibinfo{pages}{7--19} (\bibinfo{year}{2019}).

\bibitem{zheng2023}
\bibinfo{author}{Zheng, X.~Y.} \emph{et~al.}
\newblock \bibinfo{title}{Ultra-thin lithium aluminate spinel ferrite films with perpendicular magnetic anisotropy and low damping}.
\newblock \emph{\bibinfo{journal}{Nat. Commun.}} \textbf{\bibinfo{volume}{14}}, \bibinfo{pages}{4918} (\bibinfo{year}{2023}).

\bibitem{cornelissen2018three}
\bibinfo{author}{Cornelissen, L.~J.} \& \bibinfo{author}{van Wees, B.~J.}
\newblock \bibinfo{title}{Spin-current-controlled modulation of the magnon spin conductance in a three-terminal magnon transistor}.
\newblock \emph{\bibinfo{journal}{Physical Review B}} \textbf{\bibinfo{volume}{93}}, \bibinfo{pages}{020403} (\bibinfo{year}{2018}).

\bibitem{wimmer2018spin}
\bibinfo{author}{Wimmer, T.} \emph{et~al.}
\newblock \bibinfo{title}{Spin transport in a magnetic insulator with zero effective damping}.
\newblock \emph{\bibinfo{journal}{arXiv preprint arXiv:1812.01334}}  (\bibinfo{year}{2018}).

\bibitem{guckelhorn2021magnon}
\bibinfo{author}{G{"u}ckelhorn, J.} \emph{et~al.}
\newblock \bibinfo{title}{Magnon transport in {Y}$_3${F}e$_5${O}$_{12}$/{Pt} nanostructures with reduced effective magnetization}.
\newblock \emph{\bibinfo{journal}{Physical Review B}} \textbf{\bibinfo{volume}{104}}, \bibinfo{pages}{L180410} (\bibinfo{year}{2021}).

\bibitem{fulara2019spin}
\bibinfo{author}{Fulara, H.} \emph{et~al.}
\newblock \bibinfo{title}{Spin-orbit torque--driven propagating spin waves}.
\newblock \emph{\bibinfo{journal}{Science advances}} \textbf{\bibinfo{volume}{5}}, \bibinfo{pages}{eaax8467} (\bibinfo{year}{2019}).

\bibitem{collet2016generation}
\bibinfo{author}{Collet, M.} \emph{et~al.}
\newblock \bibinfo{title}{Generation of coherent spin-wave modes in yttrium iron garnet microdiscs by spin--orbit torque}.
\newblock \emph{\bibinfo{journal}{Nature communications}} \textbf{\bibinfo{volume}{7}}, \bibinfo{pages}{10377} (\bibinfo{year}{2016}).

\bibitem{Hamadeh2014SpinWaveDamping}
\bibinfo{author}{Hamadeh, A.} \emph{et~al.}
\newblock \bibinfo{title}{Full control of the spin-wave damping in a magnetic insulator using spin-orbit torque}.
\newblock \emph{\bibinfo{journal}{Physical Review Letters}} \textbf{\bibinfo{volume}{113}}, \bibinfo{pages}{197203} (\bibinfo{year}{2014}).

\bibitem{Evelt2016HighEfficiency}
\bibinfo{author}{Evelt, M.} \emph{et~al.}
\newblock \bibinfo{title}{High-efficiency control of spin-wave propagation in ultra-thin yttrium iron garnet by the spin-orbit torque}.
\newblock \emph{\bibinfo{journal}{Applied Physics Letters}} \textbf{\bibinfo{volume}{108}}, \bibinfo{pages}{172406} (\bibinfo{year}{2016}).

\bibitem{Evelt2018CoherentPropagatingMagnons}
\bibinfo{author}{Evelt, M.} \emph{et~al.}
\newblock \bibinfo{title}{Emission of coherent propagating magnons by insulator-based spin-orbit-torque oscillators}.
\newblock \emph{\bibinfo{journal}{Physical Review Applied}} \textbf{\bibinfo{volume}{10}}, \bibinfo{pages}{041002} (\bibinfo{year}{2018}).

\bibitem{Bai2024FieldFreeYIG}
\bibinfo{author}{Bai, H.} \emph{et~al.}
\newblock \bibinfo{title}{Current induced field-free switching in a magnetic insulator with enhanced spin-orbit torque}.
\newblock \emph{\bibinfo{journal}{Advanced Electronic Materials}} \textbf{\bibinfo{volume}{10}}, \bibinfo{pages}{2300785} (\bibinfo{year}{2024}).

\bibitem{Bai2025FullScaleYIG}
\bibinfo{author}{Bai, H.}, \bibinfo{author}{Li, J.}, \bibinfo{author}{Deng, X.} \emph{et~al.}
\newblock \bibinfo{title}{Full-scale field-free magnetization switching in {Y}$_3${F}e$_5${O}$_{12}$/{Pt} nanostructures for spin-orbit torque applications}.
\newblock \emph{\bibinfo{journal}{ACS Applied Nano Materials}} \textbf{\bibinfo{volume}{8}}, \bibinfo{pages}{50--55} (\bibinfo{year}{2025}).

\bibitem{Tian2020YIGReorientation}
\bibinfo{author}{Tian, Y.-Y.} \emph{et~al.}
\newblock \bibinfo{title}{Magnetization reorientation induced by spin-orbit torque in {YIG}/{Pt} bilayers}.
\newblock \emph{\bibinfo{journal}{Chinese Physics B}} \textbf{\bibinfo{volume}{29}}, \bibinfo{pages}{117504} (\bibinfo{year}{2020}).

\bibitem{Demidov2012MagneticNanoOscillations}
\bibinfo{author}{Demidov, V.~E.} \emph{et~al.}
\newblock \bibinfo{title}{Magnetic nano-oscillator driven by spin-polarized current}.
\newblock \emph{\bibinfo{journal}{Nature Materials}} \textbf{\bibinfo{volume}{11}}, \bibinfo{pages}{1028--1031} (\bibinfo{year}{2012}).

\bibitem{Tsoi1998ExcitationSpinWaves}
\bibinfo{author}{Tsoi, M.} \emph{et~al.}
\newblock \bibinfo{title}{Excitation of a magnetic multilayer by an electric current}.
\newblock \emph{\bibinfo{journal}{Physical Review Letters}} \textbf{\bibinfo{volume}{80}}, \bibinfo{pages}{4281--4284} (\bibinfo{year}{1998}).

\bibitem{Ren2025LAFOAutoOscillations}
\bibinfo{author}{Ren, H.} \emph{et~al.}
\newblock \bibinfo{title}{Electrical detection of spin-hall-induced auto-oscillations in lithium aluminate ferrite thin films}.
\newblock \emph{\bibinfo{journal}{Nano Letters}} \textbf{\bibinfo{volume}{25}}, \bibinfo{pages}{6399--6404} (\bibinfo{year}{2025}).

\bibitem{Ren2023HybridSHNO}
\bibinfo{author}{Ren, H.} \emph{et~al.}
\newblock \bibinfo{title}{Hybrid spin hall nano-oscillators based on ferromagnetic metal/ferrimagnetic insulator heterostructures}.
\newblock \emph{\bibinfo{journal}{Nature Communications}} \textbf{\bibinfo{volume}{14}}, \bibinfo{pages}{1406} (\bibinfo{year}{2023}).

\bibitem{sinova2015spin}
\bibinfo{author}{Sinova, J.}, \bibinfo{author}{Valenzuela, S.~O.}, \bibinfo{author}{Wunderlich, J.}, \bibinfo{author}{Back, C.~H.} \& \bibinfo{author}{Jungwirth, T.}
\newblock \bibinfo{title}{Spin hall effects}.
\newblock \emph{\bibinfo{journal}{Reviews of Modern Physics}} \textbf{\bibinfo{volume}{87}}, \bibinfo{pages}{1213--1260} (\bibinfo{year}{2015}).

\bibitem{maekawa2013spin}
\bibinfo{author}{Maekawa, S.}, \bibinfo{author}{Adachi, H.}, \bibinfo{author}{Uchida, K.-i.}, \bibinfo{author}{Ieda, J.} \& \bibinfo{author}{Saitoh, E.}
\newblock \bibinfo{title}{Spin current: Experimental and theoretical aspects}.
\newblock \emph{\bibinfo{journal}{Journal of the Physical Society of Japan}} \textbf{\bibinfo{volume}{82}}, \bibinfo{pages}{102002} (\bibinfo{year}{2013}).

\bibitem{uchida2010spin}
\bibinfo{author}{Uchida, K.-i.} \emph{et~al.}
\newblock \bibinfo{title}{Spin seebeck insulator}.
\newblock \emph{\bibinfo{journal}{Nature materials}} \textbf{\bibinfo{volume}{9}}, \bibinfo{pages}{894--897} (\bibinfo{year}{2010}).

\bibitem{xiao2010theory}
\bibinfo{author}{Xiao, J.}, \bibinfo{author}{Bauer, G.~E.}, \bibinfo{author}{Uchida, K.-c.}, \bibinfo{author}{Saitoh, E.} \& \bibinfo{author}{Maekawa, S.}
\newblock \bibinfo{title}{Theory of magnon-driven spin seebeck effect}.
\newblock \emph{\bibinfo{journal}{Physical Review B—Condensed Matter and Materials Physics}} \textbf{\bibinfo{volume}{81}}, \bibinfo{pages}{214418} (\bibinfo{year}{2010}).

\bibitem{Giles2015}
\bibinfo{author}{Giles, B.~L.}, \bibinfo{author}{Yang, Z.}, \bibinfo{author}{Jamison, J.~S.} \& \bibinfo{author}{Myers, R.~C.}
\newblock \bibinfo{title}{Long-range pure magnon spin diffusion observed in a nonlocal spin-seebeck geometry}.
\newblock \emph{\bibinfo{journal}{Physical Review B}} \textbf{\bibinfo{volume}{92}}, \bibinfo{pages}{224415} (\bibinfo{year}{2015}).

\bibitem{cornelissen2015long}
\bibinfo{author}{Cornelissen, L.~J.}, \bibinfo{author}{Liu, J.}, \bibinfo{author}{Duine, R.~A.}, \bibinfo{author}{Ben~Youssef, J.} \& \bibinfo{author}{van Wees, B.~J.}
\newblock \bibinfo{title}{Long-distance transport of magnon spin information in a magnetic insulator at room temperature}.
\newblock \emph{\bibinfo{journal}{Nature Physics}} \textbf{\bibinfo{volume}{11}}, \bibinfo{pages}{1022--1026} (\bibinfo{year}{2015}).

\bibitem{Uchida2008}
\bibinfo{author}{Uchida, K.} \emph{et~al.}
\newblock \bibinfo{title}{Observation of the spin seebeck effect}.
\newblock \emph{\bibinfo{journal}{Nature}} \textbf{\bibinfo{volume}{455}}, \bibinfo{pages}{778--781} (\bibinfo{year}{2008}).

\bibitem{Uchida2014}
\bibinfo{author}{Uchida, K.-i.} \emph{et~al.}
\newblock \bibinfo{title}{Longitudinal spin seebeck effect: from fundamentals to applications}.
\newblock \emph{\bibinfo{journal}{Journal of Physics: Condensed Matter}} \textbf{\bibinfo{volume}{26}}, \bibinfo{pages}{343202} (\bibinfo{year}{2014}).

\bibitem{Omahoney2023}
\bibinfo{author}{O'Mahoney, D.} \emph{et~al.}
\newblock \bibinfo{title}{Aluminum substitution in low damping epitaxial lithium ferrite films}.
\newblock \emph{\bibinfo{journal}{Applied Physics Letters}} \textbf{\bibinfo{volume}{123}}, \bibinfo{pages}{172405} (\bibinfo{year}{2023}).

\bibitem{Takana2025LowDampingLAFO}
\bibinfo{author}{Takana, L.} \emph{et~al.}
\newblock \bibinfo{title}{Low damping (111) oriented lithium aluminum ferrite thin films for spin wave applications}.
\newblock \emph{\bibinfo{journal}{Applied Physics Letters}} \textbf{\bibinfo{volume}{127}}, \bibinfo{pages}{032406} (\bibinfo{year}{2025}).

\bibitem{Alaei2025EfficientSpinTransport}
\bibinfo{author}{Alaei, S.~P.} \emph{et~al.}
\newblock \bibinfo{title}{Efficient spin transport across a disordered interface in a low damping magnetic insulator/heavy metal bilayer}.
\newblock \emph{\bibinfo{journal}{Applied Physics Letters}} \textbf{\bibinfo{volume}{127}}, \bibinfo{pages}{262403} (\bibinfo{year}{2025}).

\bibitem{tong2026direct}
\bibinfo{author}{Tong, J.} \emph{et~al.}
\newblock \bibinfo{title}{Direct observation of tunable magnons in epitaxial lithium aluminum ferrite thin films}.
\newblock \emph{\bibinfo{journal}{Nano Letters}} \textbf{\bibinfo{volume}{26}}, \bibinfo{pages}{3073--3079} (\bibinfo{year}{2026}).

\bibitem{pal2026low}
\bibinfo{author}{Pal, S.} \emph{et~al.}
\newblock \bibinfo{title}{Low-temperature spin dynamics in lithium aluminum ferrite thin films: From cubic anisotropy to tls-limited coherence}.
\newblock \emph{\bibinfo{journal}{Physical Review Materials}} \textbf{\bibinfo{volume}{10}}, \bibinfo{pages}{054406} (\bibinfo{year}{2026}).

\bibitem{Li2026NearlyIsotropic}
\bibinfo{author}{Li, Y.} \emph{et~al.}
\newblock \bibinfo{title}{Nearly isotropic magnon transport in epitaxial lithium aluminum ferrite thin films}.
\newblock \emph{\bibinfo{journal}{Applied Physics Letters}} \textbf{\bibinfo{volume}{129}}, \bibinfo{pages}{142405} (\bibinfo{year}{2026}).
\newblock \urlprefix\url{https://doi.org/10.1063/5.0341020}.

\bibitem{bender2014dynamic}
\bibinfo{author}{Bender, S.~A.}, \bibinfo{author}{Duine, R.~A.}, \bibinfo{author}{Brataas, A.} \& \bibinfo{author}{Tserkovnyak, Y.}
\newblock \bibinfo{title}{Dynamic phase diagram of dc-pumped magnon condensates}.
\newblock \emph{\bibinfo{journal}{Physical Review B}} \textbf{\bibinfo{volume}{90}}, \bibinfo{pages}{094409} (\bibinfo{year}{2014}).

\bibitem{Takei2019}
\bibinfo{author}{Takei, S.}
\newblock \bibinfo{title}{Spin transport in an electrically driven magnon gas near bose-einstein condensation: Hartree-fock-keldysh theory}.
\newblock \emph{\bibinfo{journal}{Physical Review B}} \textbf{\bibinfo{volume}{100}}, \bibinfo{pages}{134440} (\bibinfo{year}{2019}).

\bibitem{Slavin2009Nonlinear}
\bibinfo{author}{Slavin, A.} \& \bibinfo{author}{Tiberkevich, V.}
\newblock \bibinfo{title}{Nonlinear auto-oscillator theory of microwave generation by spin-polarized current}.
\newblock \emph{\bibinfo{journal}{IEEE Trans. Magn.}} \textbf{\bibinfo{volume}{45}}, \bibinfo{pages}{064047} (\bibinfo{year}{2009}).
\newblock \urlprefix\url{https://doi.org/10.1109/TMAG.2008.2009935}.

\bibitem{divinskiy2019controlled}
\bibinfo{author}{Divinskiy, B.}, \bibinfo{author}{Urazhdin, S.}, \bibinfo{author}{Demokritov, S.~O.} \& \bibinfo{author}{Demidov, V.~E.}
\newblock \bibinfo{title}{Controlled nonlinear magnetic damping in spin-hall nano-devices}.
\newblock \emph{\bibinfo{journal}{Nature communications}} \textbf{\bibinfo{volume}{10}}, \bibinfo{pages}{5211} (\bibinfo{year}{2019}).

\bibitem{Lee2022Nonlinear}
\bibinfo{author}{Lee, I.}, \bibinfo{author}{Zhang, C.}, \bibinfo{author}{Singh, S.}, \bibinfo{author}{McCullian, B.} \& \bibinfo{author}{Hammel, P.~C.}
\newblock \bibinfo{title}{Origin of nonlinear damping due to mode coupling in auto-oscillatory modes strongly driven by spin-orbit torque}.
\newblock \emph{\bibinfo{journal}{Phys. Rev. Applied}} \textbf{\bibinfo{volume}{17}}, \bibinfo{pages}{064047} (\bibinfo{year}{2022}).
\newblock \urlprefix\url{https://doi.org/10.1103/PhysRevApplied.17.064047}.

\bibitem{Dvornik2018}
\bibinfo{author}{Dvornik, M.}, \bibinfo{author}{Awad, A.~A.} \& \bibinfo{author}{{\AA}kerman, J.}
\newblock \bibinfo{title}{Origin of magnetization auto-oscillations in constriction-based spin hall nano-oscillators}.
\newblock \emph{\bibinfo{journal}{Physical Review Applied}} \textbf{\bibinfo{volume}{9}}, \bibinfo{pages}{014017} (\bibinfo{year}{2018}).

\bibitem{Vansteenkiste2014MuMax}
\bibinfo{author}{Vansteenkiste, A.} \emph{et~al.}
\newblock \bibinfo{title}{The design and verification of mumax3}.
\newblock \emph{\bibinfo{journal}{AIP Adv.}} \textbf{\bibinfo{volume}{4}}, \bibinfo{pages}{107133} (\bibinfo{year}{2014}).
\newblock \urlprefix\url{https://doi.org/10.1063/1.4899186}.

\bibitem{Tserkovnyak2002SpinPumping1}
\bibinfo{author}{Tserkovnyak, Y.}, \bibinfo{author}{Brataas, A.} \& \bibinfo{author}{Bauer, G. E.~W.}
\newblock \bibinfo{title}{Enhanced gilbert damping in thin ferromagnetic films}.
\newblock \emph{\bibinfo{journal}{Phys. Rev. Lett.}} \textbf{\bibinfo{volume}{88}}, \bibinfo{pages}{117601} (\bibinfo{year}{2002}).
\newblock \urlprefix\url{https://doi.org/ 10.1103/PhysRevLett.88.117601}.

\bibitem{Tserkovnyak2002SpinPumping2}
\bibinfo{author}{Tserkovnyak, Y.}, \bibinfo{author}{Brataas, A.}, \bibinfo{author}{Bauer, G. E.~W.} \& \bibinfo{author}{Halperin, B.~I.}
\newblock \bibinfo{title}{Nonlocal magnetization dynamics in ferromagnetic heterostructures}.
\newblock \emph{\bibinfo{journal}{Rev. Mod. Phys.}} \textbf{\bibinfo{volume}{77}}, \bibinfo{pages}{1375} (\bibinfo{year}{2005}).
\newblock \urlprefix\url{https://doi.org/10.1103/RevModPhys.77.1375}.

\end{thebibliography}

\section*{Acknowledgments}
This work was supported as part of the Center for Energy Efficient Magnonics (CEEMag), an Energy Frontier Research Center funded by the U.S. Department of Energy, Office of Science, Basic Energy Sciences at SLAC National Accelerator Laboratory under contract \# DE-AC02-76SF00515. Part of this work was performed at nano@stanford facilities RRID:SCR\_026695.

\end{document}